\documentstyle[12pt]{article}

\def\a{\alpha}
\def\b{\beta}

\def\d{\delta}

\def\m{\mu}

\def\r{\rho}
\def\s{\sigma}

\def\L{\Lambda}

\def\S{\Sigma}

\def\cu{{\cal U}}

\def\dag{\dagger}

\def\la{\left}
\def\ra{\right}
\def\pa{\partial}

\def\bar#1{\overline{#1}}

\def\Hat#1{\rlap{\kern.10em$\widehat{\phantom G}$}#1}
\def\HAt#1{\rlap{\kern.05em$\widehat{\phantom G}$}#1}

\def\cap#1{\rlap{\kern.1em$\widehat{\phantom{G\vrule height.8em}}$}#1{}}
\def\Cap#1{\rlap{\kern.05em$\widehat{\phantom{G\vrule height.8em}}$}#1{}}

\catcode`@=11
\def\underline#1{\relax\ifmmode\@@underline#1\else
        $\@@underline{\hbox{#1}}$\relax\fi}
\catcode`@=12

\def\PLB{Phys. Lett. B\ }
\def\PR{Phys. Rev.\ }

\def\PRD{Phys. Rev. D\ }

\def\NPB{Nucl. Phys. B\ }

\def\ba{\begin{array}}
\def\ea{\end{array}}
\def\be{\begin{equation}}
\def\ee{\end{equation}}
\def\bdm{\begin{displaymath}}
\def\edm{\end{displaymath}}
\def\bea{\begin{eqnarray}}
\def\eea{\end{eqnarray}}
\def\nl{\nonumber \\}

\def\by{\over}
\def\lb{\label}
\def\bl#1{(\ref{#1})}

\def\ul{\underline}
\def\ni{\noindent}

\def\vsp{\\ \vglue 0.2in}

\def\cu{{\it Department of Physics, CB 390\\
          University of Colorado, Boulder, CO 80309}}

\def\bt#1#2#3#4#5
        {\begin{titlepage}
         \centerline{\hfill {#1}}
         \centerline{\hfill {#2}}
         \centerline{\hfill {#3}}
         \begin{center}\vskip .2in
         {\large\bf {#4}}\\[.3in]
         {\bf {#5}}\\[.3in]
         {\bf ABSTRACT}\\
         \end{center}
         \begin{quotation}}
\def\et
         {\end{quotation}
          \end{titlepage}
          \setcounter{page}{2}}

\newcounter{sxn}

\def\sxnn#1{\bigskip\medskip \goodbreak \addtocounter{sxn}{1}
          \noindent{\large\bf{#1}} \nobreak \medskip}

\newcounter{axn}

\def\br{\begin{thebibliography}{abc}}
\def\er{\end{thebibliography}}
\def\rf{\bibitem}
\def\fr#1{\cite{#1}}
\begin{document}

\bt{COLO-HEP/325}{hep-th/9310094}{October 1993}{Composite gauge field
models with broken symmetries}{B. S. Balakrishna\footnote{Email:
bala@haggis.colorado.edu} and K. T. Mahanthappa\footnote{Email:
ktm@verb.colorado.edu} \vsp \cu}

We present a generalization of the non-Abelian version of the
$CP^{N-1}$ models (also known as Grassmannian models) that involve
composite gauge fields to accommodate partial breaking of the
non-Abelian gauge symmetry. For this to be possible, in most cases,
the constituent fields need to belong to an anomaly free complex
representation. Symmetry is broken dynamically for large $N$ primarily
by a naturally generated composite scalar which simulates Higgs
mechanism. In the example studied in some detail, the gauge group
SO(10) gets broken down to subgroups like SU(5) or SU(5)$\times$U(1).
\\
\\
\ni PACS nos: 11.15.Ex, 12.50.Fk

\et

The concept of composite gauge fields is attractive because of a
possible reduction in the number of fields, economy in the number of
parameters and the expectation of softer ultraviolet divergences.
They have a long history\fr{old,cpn,gms,pal}. They appear in
$CP^{N-1}$ models\fr{cpn} and their non-Abelian generalizations called
Grassmannian models(GMs)\cite{gms} as composites of bosonic or
fermionic\fr{pal} constituents. These models have been studied in the
large $N$ limit, and it is found that the non-Abelian symmetry is
either exact or completely broken. If such composite models are to be
candidates for a physical theory, we do need a version in which the
symmetry is partially broken. The purpose of this note is to present
one such version. It is a generalization of the GMs made possible due
to a modification of the constraint equation by a natural inclusion of
a composite scalar in the adjoint representation of the gauge group
simulating Higgs mechanism. It is interesting to note here that the
agent of symmetry breaking in grand unified theories is usually a
Higgs scalar in the adjoint representation. In the following, we
confine ourselves to bosonic constituent fields. In the case of the
gauge group SO(10) studied in some detail, there exists a rich phase
structure with symmetry breaking to subgroups like SU(5) or
SU(5)$\times$U(1).

The Grassmannian model\fr{gms} is based on $N$ scalars in the
fundamental representation of the gauge group, represented
collectively by a $M\times N$ matrix $Z$ with the elements $Z_{\a i}$.
The column index $i$ is an internal index or a flavor index. The row
index $\a$ is the gauge index associated with a non-Abelian gauge
symmetry which in the present case is U($M$). In other words, the
action of $U(M)$ on $Z$ is from the left, $Z\to UZ$. The case of the
$CP^{N-1}$ model can be recovered from GM by setting $M=1$. The model
is defined under the constraint $ZZ^\dag=I_M$ where $I_M$ is an
identity matrix of order $M$. The Lagrangian written in terms of an
auxiliary field $A_\m=iZ\pa_\m Z^\dag$ reads
\be
L ~=~ \b N{\rm tr}\la[D_\m Z(D_\m Z)^\dag\ra],
\lb{l1}
\ee
where $D_\m Z$ is the covariant derivative $(\pa_\m-iA_\m)Z$; $\b$ is
an inverse coupling constant. An overall multiplicative factor $N$ is
introduced for convenience in the $1/N$ expansion. It is easy to
verify that the composite field $A_\m$ transforms as a gauge field
under the local transformation $Z\to UZ$, thanks to the constraint. It
is expected to become dynamical and hence a genuine gauge field after
quantum corrections. This can be seen in the $1/N$ expansion, for
instance.

The phase structure of these models has been studied in the literature
for large $N$\cite{gms}. It is determined by a critical coupling $\b_c$
given by
\be
\b_c ~=~ \int{d^4k\by(2\pi)^4}{1\by k^2},
\ee
where, as in the rest of this note, we have suppressed the dependence
on the momentum cutoff. There are two phases. For $\b>\b_c$, there is
the broken phase where $Z$ has an expectation value. This breaks
$U(M)$ completely and all the gauge bosons become massive. For
$\b<\b_c$, we have the unbroken phase where the gauge symmetry is
unbroken and the gauge bosons are massless. In other words, {\it the
gauge group is either completely broken or not broken at all}.
Apparently there are no phases, at least for large $N$, where a
partial breaking of the gauge group is possible. These features follow
as a special case of our general discussion below. To obtain a richer
phase structure, we invoke a generalization of these models and study
them for large $N$ in the following.

A straightforward approach is to retain the constraint $ZZ^\dag=I_M$
as such, but to gauge only a subgroup of U($M$). This, however, does
not alter the the symmetry breaking patterns. A more interesting
generalization occurs when the constraint itself is modified. Here we
take the $Z$ fields to be in any representation of the gauge group,
say $R$, of dimension $M$ and multiplicity $N$. The transformation
matrix $U$ is now in $R$ acting on the matrix $Z$ as before, $Z\to
UZ$. We look for a Lagrangian that is of the form \bl{l1}. Note that
the part of the Lagrangian quadratic in $A_\m=A_\m^aT_a$ is
proportional to $A_\m^aA_\m^b{\rm tr}(T_aT_bZZ^\dag)$, where $T_a$'s
are the generators of the Lie algebra being gauged. In the GM, the
constraint $ZZ^\dag=I_M$ is responsible for rendering it quadratic in
$A_\m$ alone. Now, more generally, we achieve the same goal by
imposing the following constraint instead:
\be
{\rm tr}(T_{ab}ZZ^\dag) ~=~ l\d_{ab}.
\lb{con}
\ee
Here $T_{ab}=(T_aT_b+T_bT_a)/2$ and $l$ is the index of the
representation $R$ defined by ${\rm tr}(T_aT_b)=l\d_{ab}$. Note that
this new constraint also respects the gauge symmetry. Using it, it is
easy to obtain an expression for the composite field $A_\m$,
\be
A_\m^a ~=~ {i\by 2l}{\rm tr}\la[T_a(Z\pa_\m Z^\dag-\pa_\m
ZZ^\dag)\ra],
\ee
and an expression for the Lagrangian in terms of the $Z$ fields alone,
\be
L ~=~ \b N{\rm tr}(\pa_\m Z\pa_\m Z^\dag) + {1\by 4l}\b N\la\{{\rm
tr}\la[T_a(Z\pa_\m Z^\dag-\pa_\m ZZ^\dag)\ra]\ra\}^2.
\ee
This derivation should ensure gauge invariance and this is easily seen
to be the case.

Our earlier constraint $ZZ^\dag=I_M$ clearly solves \bl{con}. But,
there are cases where this is not the only solution. In other words,
the constraint \bl{con} is in some cases weaker than the earlier one.
To see this, introduce a hermitian matrix $W$ by $ZZ^\dag=I_M+W$ and
observe that the constraint is equivalent to looking for a solution of
${\rm tr}(T_{ab}W)=0$. Given a $W$ that leads to a positive
semidefinite $ZZ^\dag$, $Z$ is solvable generally as
$Z=(I_M+W)^{1/2}Z_0$ for some $Z_0$ obeying $Z_0Z_0^\dag=I_M$. The
earlier constraint corresponds to the trivial solution $W=0$. That
there exist cases where $W$ is nontrivial can be seen as follows. Make
the ansatz that $W$ is in the Lie algebra itself, i.e., $W=W_aT_a$.
Now, the constraint ${\rm tr}(T_{ab}W)=0$ simply states that the ABJ
anomaly associated with the representation $R$ should vanish. That
there exist anomaly free representations is well known. It is easy to
observe that the representation $R$ should be complex for an ansatz of
this type to solve the constraint\cite{chi}. An example is the spinor
representation 16 of the gauge group SO(10) that is well known to be
anomaly free. Our ansatz gives a solution for the composite scalar $W$
that is in the adjoint representation 45. This is interesting, for it
is known that an adjoint scalar is a promising candidate to break
SO(10) to SU(5). Its appearance here is quite unexpected.

In some instances, the above ansatz gives the most general solution.
This is the case for the SO(10) example mentioned above. This can be
seen from representation theory, viewing $W$ to belong to
$16\times\bar{16}=1+45+210$. It can be shown that the constraint,
involving a symmetric product of the generators of the gauge group,
suppresses the representations 1 and 210 from $W$ leaving only the 45
as claimed. Note, however, that the ansatz does not always give the
most general solution. For instance, if one were to pick a
sufficiently large representation of the gauge group for $R$, one will
easily end up with more representations that remain unsuppressed in
$W$. Representation theory should still be applicable to solve for $W$
in general.

Let us call the models as type one models when $W$ is identically zero
and our new constraint reduces to the old one. They are closer to the
GMs discussed in the beginning, or rather to their generalizations
involving subgroups of U($M$). The other models where $W$ can be
nonzero is referred to as type two models. Models based on a reducible
$R$ are quite generally of type two. This is because the constraint
does not determine some components of $W$, for instance those
connecting different subrepresentations\cite{qns}. Type one models are
thus necessarily based on irreducible $R$'s.

The constraint \bl{con} can be incorporated into the Lagrangian with
the help of a Lagrangian multiplier $\S=\S_{ab}T_{ab}$, a $M\times M$
matrix. The result is
\be
L ~=~ \b N{\rm tr}\la[D_\m Z(D_\m Z)^\dag + \S ZZ^\dag - \S\ra].
\lb{l2}
\ee
To understand symmetry breaking, and hence to identify the various
phases, we need to obtain the effective potential. We will do this for
large $N$. Because $A_\m$ is not expected to pick up any expectation
value, it can be set to zero. The contribution to the effective
potential coming from the Lagrangian \bl{l2} is obtained by dropping
the derivative terms. Because $1/N$ appears in the Lagrangian like the
Planck's constant, the quantum corrections to this contribution are
expected to be suppressed by a factor $1/N$. But there are $N$
representations contributing equally and this can offset the $1/N$
suppression. The result is that for large $N$ the effective potential
for the $Z$ and $\S$ fields obtained by integrating away the $Z$
fluctuations carries a correction
\be
N\int{d^4k\by(2\pi)^4}{\rm tr}~{\rm ln}\la(k^2I_M+\S\ra).
\ee
The total effective potential is thus
\be
V_{\rm eff} ~=~ \b N{\rm tr}\la(\S ZZ^\dag-\S\ra) +
N\int{d^4k\by(2\pi)^4}{\rm tr}~{\rm ln}\la(k^2I_M+\S\ra).
\ee
To determine the various phases, we need to extremize this potential.
The resulting saddle point equations are $\S Z=0$ obtained by varying
$Z^\dag$ and
\be
\b{\rm tr}\la[T_{ab}(ZZ^\dag-I_M)\ra] + \int{d^4k\by(2\pi)^4}{\rm
tr}\la(T_{ab}{1\by k^2I_M+\S}\ra) ~=~ 0
\lb{tte}
\ee
coming from varying $\S$. We now look for solutions to this system of
equations.

For type one models, the traces and the $T_{ab}$'s can be dropped and
Eq. \bl{tte} becomes equivalent to that of the GM. The solutions lead
to two phases\fr{gms}. There is the broken phase for $\b>\b_c$ with
$ZZ^\dag=(1-\b_c/\b)I_M$ and $\S=0$, and the unbroken phase for
$\b<\b_c$ with $Z=0$ and $\S=\s I_M$. Here $\s$ is given by
\be
\int{d^4k\by(2\pi)^4}{1\by k^2+\s} ~=~ \b.
\ee
We have to ensure that the solution for $\S$ is consistent with its
definition $\S=\S_{ab}T_{ab}$. $\S=\s I_M$ is acceptable since it is
equivalent to $\S_{ab}=\s\d_{ab}/C_2(R)$, where $C_2(R)$ is the second
Casimir invariant of the irreducible representation $R$ defined by
$T_aT_a=C_2(R)I_M$.

For type two models, Eq. \bl{tte} can still be reduced to resemble
that of the GM, but with a matrix $W$ satisfying ${\rm tr}(T_{ab}W)=0$
on the r.h.s,
\be
\b\la(ZZ^\dag -I_M\ra) + \int{d^4k\by(2\pi)^4}{1\by k^2I_M+\S} ~=~ \b W.
\lb{rsw}
\ee
The factor $\b$ on the r.h.s makes this equation agree with
$ZZ^\dag=I_M+W$ at the level of expectation values. For the unbroken
case of $\b<\b_c$, we have the previous solution $\S=\s I_M$, $Z$ and
$W$ being zero. $\S=\s I_M$ is acceptable for irreducible $R$'s as it
follows from $\S_{ab}=\s\d_{ab}/C_2(R)$. The case of reducible $R$'s
can be handled starting with the ansatz $\S_{ab}=\s\d_{ab}$ suitably
choosing $W$. For the completely broken case of $\b>\b_c$, the
solution is $ZZ^\dag=(1-\b_c/\b)I_M+W$ where $W$ is arbitrary to the
extent that $ZZ^\dag$ is positive semidefinite.

These solutions are not a priori the most general ones. We use our
SO(10) example to look for more solutions. The solution we are looking
for is intended to break SO(10) to SU(5) or to SU(5)$\times$ U(1). Our
following attempt is not successful, but illustrates the way more
solutions could arise. Besides, it is an exercise that will be useful
to our discussion later. Note that SO(10) has a maximal subgroup
SU(5)$\times$U(1). The representation 16 of SO(10) transforms under
SU(5) as $10+\bar{5}+1$. The U(1) charges for 10, $\bar{5}$ and 1 are
proportional to 1,$-3$ and 5 respectively. Consider first the case
$Z=0$. In order to have symmetry breaking to SU(5)$\times$U(1), we now
take $\S_{ab}=\s\d_{ab}$ along the SU(5) directions, $\r$ along the
U(1) and zero otherwise\cite{sym}. Note that $a$ or $b$ index runs
over the adjoint representation 45 of SO(10) that under SU(5)
decomposes to $24+10+\bar{10}+1$. Our ansatz for $\S_{ab}$ corresponds
to having it nonzero for $(a,b)$ along $(24,24)$ and $(1,1)$. One
could have it nonzero along $(10,\bar{10})$ and $(\bar{10},10)$ as
well, but it turns out that this can be absorbed into $\s$ and $\r$.
This means that the $\S$ matrix is diagonal with values
$C_2(10)\s+\r$, $C_2(\bar{5})\s+9\r$ and $25\r$ along the
representations 10, $\bar{5}$ and 1 respectively. With
$C_2(10)/C_2(\bar{5})=3/2$ and a suitable scaling of $\s$, we may take
them to $3\s+\r$, $2\s+9\r$ and $25\r$. The $W$ matrix is taken to be
along the U(1) direction; in other words, it is diagonal with values
$w$, $-3w$ and $5w$. It is now straightforward to write down the
saddle point equations,
\bea
-\b + \int{d^4k\by(2\pi)^4}{1\by k^2+3\s+\r} &=& \b w, \nl -\b +
\int{d^4k\by(2\pi)^4}{1\by k^2+2\s+9\r} &=& -3\b w, \nl -\b +
\int{d^4k\by(2\pi)^4}{1\by k^2+25\r} &=& 5\b w.
\lb{so1}
\eea
We find no solutions to these equations other than the one already
discussed wherein $\S$ is proportional to identity and $W=0$. Note
that the $\S$ eigenvalues $3\s+\r$, $2\s+9\r$ and $25\r$ are either in
the ascending order or in the descending order (or equal), and this
makes it difficult to obtain the alternating signs on the r.h.s of the
above equations. Allowing for nonvanishing $Z$ does not improve the
situation. This is due to the requirement that $ZZ^\dag$ and $\S$ have
nonnegative eigenvalues, and due to the equation $\S ZZ^\dag=0$ that
requires at least $3\s+\r$ or $25\r$ to vanish to allow for a nonzero
eigenvalue of $ZZ^\dag$. We have also examined the case of $E_6$ where
$R$ stands for the representation 27 for possible breaking to SO(10)
or SO(10)$\times$U(1) and find no solutions.

Perhaps this is illustrative of a generic phenomenon or suggestive of
the need to look at larger representations that might lead to more
solutions. These examples, though not successful in a partial breaking
of the gauge symmetry, will be useful to us below where a potential is
introduced leading to a rich phase structure. The adjoint scalar $W$,
that has not played any significant role so far, is going to play a
major one in the presence of a potential.

One could investigate models with potential terms that are polynomials
in $Z$ and $Z^\dag$. There is no simple way to do this in the
canonical GM without spoiling the local and global symmetries and the
constraint equation. However, the unexpected appearance of an adjoint
scalar $W$ helps us to construct suitable potentials, and this allows
for a partial breaking of the gauge symmetry. Let us keep the
potential quite general to begin with, $\b N{\rm tr}V(ZZ^\dag)$, where
$V(\cdot)$ is some polynomial in its argument. We expect this to be a
nontrivial extension only in the case of type two models, since for
type one the constraint $ZZ^\dag=I_M$ reduces it to the addition of a
constant. It is convenient to introduce a composite field variable $X$
for $ZZ^\dag$ and write the potential as $\b N{\rm tr}V(X)$. The
requirement $X=ZZ^\dag$ can be incorporated with the help of a
Lagrange multiplier $Y$, adding a term $\b N{\rm tr}(YZZ^\dag-YX)$ to
the potential. As before, the constraint ${\rm
tr}(T_{ab}ZZ^\dag)=l\d_{ab}$ can be accommodated with the help of a
Lagrange multiplier $\S=\S_{ab}T_{ab}$. Its effect is, as we know, to
add a term $\b N{\rm tr}(\S ZZ^\dag-\S)$ to the potential. After
translating $Y$ to $Y-\S$ for convenience, the total Lagrangian looks
like
\be
L ~=~ \b N{\rm tr}\la[D_\m Z(D_\m Z)^\dag +V(X) +YZZ^\dag -YX +\S X
-\S\ra].
\ee
The large $N$ effective potential is now computable,
\be
V_{\rm eff} ~=~ \b N{\rm tr}\la[V(X)+YZZ^\dag-YX+\S X-\S\ra] +
N\int{d^4k\by(2\pi)^4}{\rm tr}~{\rm ln}\la(k^2I_M+Y\ra).
\ee
The saddle point equations are obtained by extremizing this potential.
Varying $X$ gives $Y=\S+V'(X)$ where a prime denotes differentiation
with respect to the argument. Varying $\S$ gives ${\rm
tr}[T_{ab}(X-I_M)]=0$. As before, one may look for a solution to this
in the form $X=I_M+W$ where $W$ satisfies ${\rm tr}(T_{ab}W)=0$. These
solutions simply determine $X$ and $Y$. Varying $Y$ and using these
results one gets
\be
\b\la(ZZ^\dag -I_M\ra) + \int{d^4k\by(2\pi)^4}{1\by
k^2I_M+\S+V'(I_M+W)} ~=~ \b W.
\ee
This is to be supplemented with $YZ=\la[\S+V'(I_M+W)\ra]Z=0$ obtained
by varying $Z^\dag$. This system of equations resembles the one
obtained earlier (see Eq. \bl{rsw}) with $\S$ replaced by
$\S+V'(I_M+W)$. The presence of $V'(I_M+W)$, however, is suggestive of
a different phase structure.

For type one models $W=0$, and $V'(I_M+W)$ just adds a constant to
$\S$. This can be absorbed into $\S$ because these models, being based
on an irreducible $R$, allow for the addition of a term proportional
to identity to $\S$. As expected earlier, this is a trivial extension.
However, this is not the case for type two models and we expect a rich
phase structure.

As an example, consider a sixth order potential in $Z,Z^\dag$ that
leads to $V'(I_M+W) = aI_M+bW+cW^2$ for some parameters $a$, $b$ and
$c$. For a model based on an irreducible $R$, the term $aI_M$ can be
absorbed into $\S$ as we have already noted. When $W$ is an adjoint
variable, the term $cW^2$ can also be absorbed into $\S$. In the
SO(10) example, these correspond to translating $\s$ to $\s-8a/25$ and
$\r$ to $\r-a/25-cw^2$. With this taken care of, the set of equations
to be solved in our SO(10) example is (for $Z=0$)
\bea
-\b + \int{d^4k\by(2\pi)^4}{1\by k^2+3\s+\r+bw} &=& \b w, \nl -\b +
\int{d^4k\by(2\pi)^4}{1\by k^2+2\s+9\r-3bw} &=& -3\b w, \nl -\b +
\int{d^4k\by(2\pi)^4}{1\by k^2+25\r+5bw} &=& 5\b w.
\lb{feq}
\eea
Note the addition of $bw$ terms in the denominators compared to Eq.
\bl{so1}. These equations do have solutions for a range of parameters.
This can be seen by treating $x=3\s+\r+bw$ and $y=2\s+9\r-3bw$ as
independent variables to determine $\b$ and $w$ from the first two
equations, and $b$ from the last one. Both $x$ and $y$ should remain
positive (or zero) to keep the momentum integrals well defined.
Solutions are noted to exist in a certain domain of $x$ and $y$ giving
rise to a range for the parameters. This happens for $\b<\b_c$ and
$b<0$\fr{hig}. The results of our numerical investigation is presented
in Fig. 1. There are in fact two solutions for a given $\b$ and $b$ in
the region between the curves (a) and (b), and one solution between
the curves (b) and (c). In other words, one of the solutions extends
from curve (a) to curve (b) while the other from curve (a) to curve
(c). The symmetry breaking involved here is from SO(10) to
SU(5)$\times$U(1) as noted before.

One also finds solutions for a nonzero $Z$. Consider giving an
expectation value $v^2$ for $ZZ^\dag$ along the singlet in the
decomposition $16=10+\bar{5}+1$. In this case, the last equation above
is modified to
\be
\b v^2 -\b + \b_c ~=~ 5\b w,
\lb{leq}
\ee
where we have set $z=25\r+5bw$ to zero to satisfy $YZ=0$. That there
exist solutions yielding a positive $v^2$ for a range of parameters
can again be seen by treating $x$ and $y$ as independent variables to
determine the others. These solutions also require $\b<\b_c$ and a
negative $b$. The region of the parameter space covered by them (one
solution for a given $\b$ and $b$) is that in between the dashed
curves (b) and (d) of Fig. 1. The symmetry is broken from SO(10) to
SU(5). There are other possibilities. Giving an expectation value for
$ZZ^\dag$ along $\bar{5}$ (instead of the singlet) also leads to a
solution which falls above the curve (c). Solutions are also noted to
exist for a nonzero $ZZ^\dag$ along 10 and 1 (extending above curve
(d)), or 10 and $\bar{5}$ (extending beyond that of $\bar{5}$). All of
these, however, break the gauge group completely.

What we have in the end is a two sheeted cover of the parameter space
above the critical curve (a) in Fig. 1. One of them (call it the upper
sheet) is through the solid curves while the other one (call it the
lower sheet) is through the dashed curves. They meet along curve (a).
There is of course one more sheet (call it the top sheet) for the
solutions of our earlier case of the unbroken gauge group covering all
of the parameter space for $\b<\b_c$. This too meets the other two
sheets along curve (a). For every point on any one of the sheets,
there is a solution.

There could be more solutions. For instance, there is the possibility
that a solution breaking SO(10) to SU(3)$\times$SU(2)$\times$U(1)
(perhaps, with an additional U(1)) exists. The number of variables and
the number of equations at least matches, each being six. This
suggests the need to look at all the possibilities, including
SO(10)$\to$SU(4), SU(4)$\times$U(1), etc. The system of equations are
complicated to handle, and we defer their study to the future.

Now, consider all the solutions for $\b<\b_c$, i.e., both unbroken and
partially broken ones. Which solution is preferred is of course
determined by the effective potential. For this, one needs to compute
$V_{\rm eff}$ for all the solutions and pick the one (or more) that
has the the lowest value. This is not an easy task given the number of
possibilities involved, and hence we will be content with doing this
numerically for the solutions found above. Our purpose here is to show
that there exists a range of the parameter space that prefers a
partial breaking of the gauge group.

It is straightforward to obtain the following expression for the
effective potential at a saddle point:
\be
V_{\rm eff} ~=~ -\b N{\rm tr}\la(\S+bW^2/2\ra) +
N\int{d^4k\by(2\pi)^4}{\rm tr}~{\rm ln}\la(k^2I_M+\S+bW\ra).
\lb{vsp}
\ee
As usual, we are concerned with an adjoint $W$ for an irreducible $R$,
with the SO(10) example in mind. We have again absorbed the terms $aI$
and $cW^2$ into $\S$ for convenience. Fig. 2 is a plot of this
effective potential for the three sheets involved (we have chosen a
path in the plane of Fig. 1 crossing all the curves suitably fixing
$z$ in the upper sheet and $y$ in the lower sheet). The uppermost
curve is for the top sheet, the middle one is for the upper sheet and
the lowermost one is for the lower sheet. Note that the lower sheet
ends up always having the lowest potential. In other words, for
$\b<\b_c$ but not close to it, a partial breaking of the gauge group
is preferred over the unbroken case.

What is remarkable of our exercise in SO(10) is that a set of
equations governed by only two parameters ($\b$ and $b$) gives rise to
a rich set of solutions with interesting symmetry breaking patterns.
Apparently, there exist regions of the parameter space where SO(10)
breaking to SU(5), or to SU(5)$\times$U(1), or perhaps to some other
groups is possible. This example achieves our goal of constructing an
induced gauge theory with composite gauge bosons having partial
symmetry breaking.

There are some issues that have not been addressed here. The kinetic
terms for the gauge fields need to be computed to determe the gauge
coupling constant and the gauge boson masses. The kinetic term for the
adjoint scalar when computed will determine its properties. The fate
of the global symmetry and the number of Goldstone bosons that could
emerge from its breaking need to be worked out. Also, the possible
existence of other breakings of SO(10) needs to be explored. Results
of these investigations will be published elsewhere. There is of
course the issue of renormalizability. It is interesting to note in
this connection that renormalizable models of composite gauge fields
have been constructed in ref. \cite{pal}.

This work is supported in part by U. S. Department of Energy, Grant
No. DEFG-ER91-40672.

\br
\rf{old}
J. D. Bjorken, Ann. Phys. (N.Y.) 24, 174 (1963); G. S. Guralnik, \PR
136, 1404 (1964); T. Eguchi, \PRD 14, 2755 (1976); H. Terezawa, Y.
Chikashige and K. Akama, {\it ibid.} 15, 480 (1977); C. Bender, F.
Cooper and G. Guralnik, Ann. Phys. (N.Y.) 109, 165 (1977); K. Shizuya,
\PRD 21, 2237 (1980). D. Amati, R. Barbieri, A. C. Davis and G.
Veneziano, \PLB 102, 408 (1981); A. Hasenfratz and P. Hasenfratz, \PLB
297, 166 (1992).
\rf{cpn}
H. Eichenherr, \NPB 146, 215 (1978); A. D'Adda, P. DiVecchia and M.
Lusher, {\it ibid.} 146, 63 (1978); 152, 125 (1979); E. Witten, {\it
ibid.} 149, 285 (1979).
\rf{gms}
E. Gava. R. Jengo and C. Omero, \PLB 81, 187 (1979); \NPB 158, 381
(1979); E. Bre\'{z}in, S. Hikami and J. Zinn-Justin, {\it ibid.} 165,
528 (1980); S. Duane, {\it ibid.} 168, 32 (1980). We do not include
explicit symmetry breaking terms as in Bre\'{z}in et. al. Further, our
discussion is carried out in four dimensional Euclidean space-time.
\rf{pal}
F. Palumbo, \PRD 48, 1917 (1993).
\rf{chi}
Interestingly, for a consistent chiral theory relevant to nature, the
fermions are required to belong to an anomaly free complex
representation. Here we encounter the same requirement, though in the
bosonic version.
\rf{qns}
The case of subrepresentations being all alike can be viewed as type
one with all its multiplicity dumped to $N$.
\rf{sym}
Note that for $Z=0$, the unbroken symmetry corresponds to those
generators that commute with the ansatz for $\S$, and this happens to
be SU(5)$\times$U(1). A nonvanishing $Z$ along the singlet in the
decomposition $16=10+\bar{5}+1$ would break the U(1) as well leaving
only SU(5) unbroken.
\rf{hig}
A negative $b$ corresponds to the adjoint scalar $W$ having a negative
mass squared in its potential, as in the Higgs mechanism.
\er

\sxnn{Figure captions}

\ni\ul{Fig. 1:} The regions of parameter space of $\b$ and $b$ that
yield solutions to Eq. \bl{feq}, $\L$ being the momentum cutoff. See
the explanations after Eqs. \bl{feq} and \bl{leq} in the text.

\ni\ul{Fig. 2:} A plot of the effective potential (with its zero
appropriately chosen) versus $\b$ for some path that crosses all the
curves in Fig. 1. The crossings are denoted by (a), (b), (c) and (d).
The uppermost curve corresponds to the unbroken case and the lower two
correspond to symmetry breaking as discussed in the text after Eq.
\bl{vsp}.

\vfill\eject

\let\picnaturalsize=N
\def\picsize{5.5in}
\def\picfilename{ig.eps}
\ifx\nopictures Y\else{\ifx\epsfloaded Y\else\input epsf \fi
\let\epsfloaded=Y
\centerline{\ifx\picnaturalsize N\epsfxsize \picsize\fi
\epsfbox{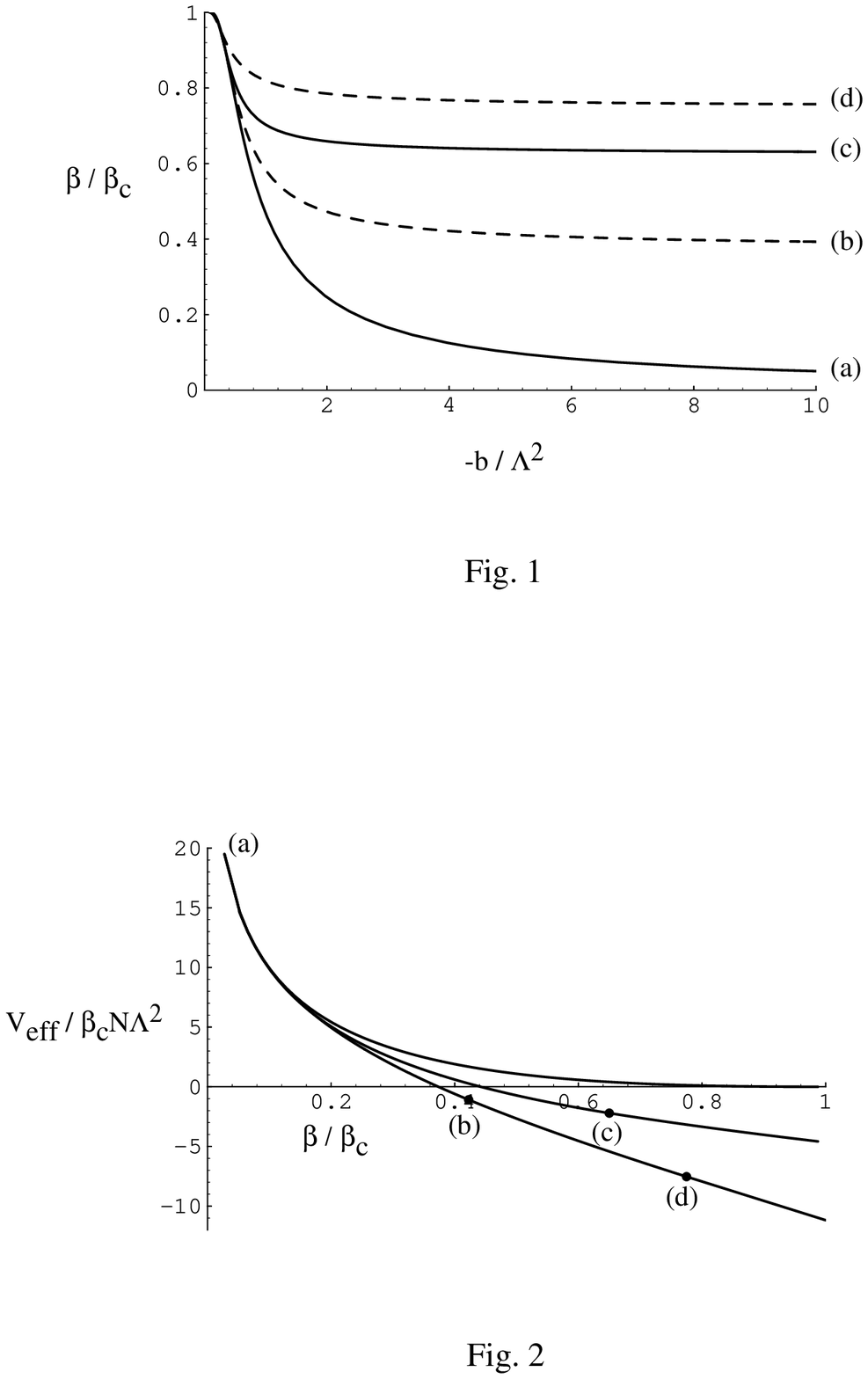}}}\fi

\end{document}